\documentclass[epj,final]{svjour}

\usepackage{color}

\usepackage{graphicx}
\usepackage{amsmath,amssymb}
\usepackage{psfrag}

\usepackage[colorlinks=true,linkcolor=blue,citecolor=blue,urlcolor=blue]{hyperref}

\newcommand{\dd}{{\rm d}}

\newcommand{\kx}{K_{{\rm air}\,x}}
\newcommand{\ky}{K_{{\rm air}\,y}}
\newcommand{\oml}{\Omega_{\rm air}}
\newcommand{\om}{\omega}
\newcommand{\kxp}{k_{x}}
\newcommand{\kyp}{k_{y}}
\newcommand{\kxg}{K_{x}}
\newcommand{\kyg}{K_{y}}
\newcommand{\omp}{\omega}
\newcommand{\omg}{\Omega}

\newcommand{\kg}{{K}}

\newcommand{\kxb}{k_{{\rm air}\,x}}
\newcommand{\kyb}{k_{{\rm air}\,y}}
\newcommand{\omb}{\omega_{\rm air}}

\newcommand{\omlow}{\om_{\rm t}}
\newcommand{\ommin}{\om_{\rm min}}
\newcommand{\ommax}{\om_{\rm max}}

\newcommand{\omgap}{\Omega_{\rm gap}}

\newcommand{\omgone}{\Omega_{1}}
\newcommand{\omgtwo}{\Omega_{2}}
\newcommand{\omgthree}{\Omega_{3}}

\newcommand{\omlone}{\Omega_{{\rm air}\,1}}
\newcommand{\omltwo}{\Omega_{{\rm air}\,2}}

\newcommand{\ompone}{\om_{1}}
\newcommand{\omptwo}{\om_{2}}

\newcommand{\thetal}{\theta_{\rm air}}
\newcommand{\thetagl}{{\theta}}

\newcommand{\omlaser}{\Omega_p}

\title{Kinematic study of the effect of dispersion in quantum vacuum emission from strong laser pulses}
\titlerunning{Kinematics study of quantum vacuum emission from strong laser pulses} 

\author{S. Finazzi\thanks{E-mail:~\href{mailto:finazzi@science.unitn.it}{finazzi@science.unitn.it}} \and I. Carusotto}

\institute{INO-CNR BEC Center and Dipartimento di Fisica, Universit\`a di Trento, via Sommarive 14, 38123 Povo-Trento, Italy}

\date{}
\abstract{
A strong light pulse propagating in a nonlinear medium causes an effective change in the local refractive index. With a suitable tuning of the pulse velocity, the leading and trailing edge of the pulse were predicted to behave as analogue black and white horizons in the limit of a dispersionless medium. In this paper, we study a more realistic situation where the frequency dispersion of the medium is fully taken into account. As soon as positive frequency modes with negative norm are present in the comoving frame, spontaneous emission of quantum vacuum radiation is expected to arise independently of the presence of horizons. We finally investigate the kinematic constraints put on the emission and we show that the optimal directions to observe Hawking-like emission form a narrow angle with the direction of propagation of the pulse.
} 

\begin{document}

\maketitle

\section{Introduction}

Hawking radiation~\cite{hawking,hawking75} is the quantum production of particles from vacuum fluctuations due to the presence of a black hole horizon in a curved stationary geometry. However, this prediction does not reside on the peculiar dynamical features of a general-relativistic spacetime, but only on the kinematic properties of a quantum field living in a curved spacetime.
Following the pioneering work~\cite{unruh}, analogous quantum vacuum emission phenomena have been predicted to occur in other physical systems, provided that a quantum field propagates on an effectively curved spacetime endowed with a horizon (see~\cite{lr} for a complete review on analogue models of gravity).

In particular, optical systems provide promising experimental analogues of curved spacetimes~\cite{ulf_science,transfoptics}, because they are quite simple to handle and the technology to perform the required measurements is well developed.
In fact, the first claim of analogue Hawking radiation in a laboratory has been recently reported in~\cite{faccioexp,faccioexplong}, where the analogue horizons are created by sending a strong laser pulse through a nonlinear dielectric medium with an appropriate velocity. The claimed observation of optical Hawking radiation was obtained by looking perpendicularly to the direction of propagation of the pulse, so to eliminate other spurious effect.
Unfortunately, this result is still considered as controversial by some authors, who recently raised a few issues~\cite{comment,reply}. Nonetheless, no convincing alternative explanation of the experimental observations has been found yet. It is worth mentioning some attempts to interpret them as a dynamical Casimir emission~\cite{angus} and/or the optical analogue of cosmological particle creation~\cite{newunruh}.

In this paper, we extend the existing models for analogue Hawking radiation produced by a strong pulse propagating through a nonlinear optical medium, by considering the kinematic effect of the frequency dispersion on the emission process. This allows to put constraints on the observable features of the emission and suggests the most suitable experimental conditions for its observation.

Focusing our attention on a fused silica glass as the nonlinear medium, we first study a one-dimensional system. 
In a nondispersive stationary medium, the necessary and sufficient condition to trigger particle creation {\it \`a la} Hawking would be the presence of analogue horizons and the spectrum of the spontaneously emitted particles would be Planckian at a temperature proportional to the surface gravity of the horizon~\cite{faccioprd}.
In dispersive media such as fused silica, the situation is kinematically more complicate, but quantum vacuum emission remains still possible: while horizons are no longer necessary for the emission, a sufficient condition is the presence of positive-norm modes
with negative frequency in the reference frame comoving with the pulse. However, in this case, there is no {\it a priori} reason for the spectrum to be thermal.

In the second part of the paper, we study the propagation of photons in a three-dimensional system with axial symmetry. 
In particular, we investigate how the kinematical properties of the emission in the comoving frame transfer into the observable radiation in the laboratory and we determine which components of the quantum vacuum radiation are able to actually exit the glass reaching the detector.

\section{Mode analysis in one-dimensional media}

We consider a laser pulse with frequency $\omlaser$, propagating with velocity $V$ in a parallelepiped of glass.
If the pulse is strong enough, the refractive index experienced by a probe field increases~\cite{butchercotter} proportionally to the laser intensity $I$:
\begin{equation}
 n(I)=n_0+\delta n(I)=n_0+n_2\,I,
\end{equation}
where $n_2\approx3\times10^{-16}\,\mbox{cm}^2/\mbox{W}$.
For a realistic value of $I\approx3\times10^{12}\,\mbox{W}/\mbox{cm}^2$, $\delta n\approx 0.001$~\cite{faccioexp,faccioexplong}.

Consequently, the velocity of propagation $v_{\rm in}$ of the probe field inside the pulse is smaller than outside ($v_{\rm out}$). When neglecting the effects of dispersion and if the parameters are arranged such that $v_{\rm in}<V<v_{\rm out}$, photons propagating inside the pulse cannot cross the leading edge, but they are all dragged towards the trailing one.
The leading (trailing) edge is then the analogue of a black (white) horizon, as seen by an observer outside the pulse~\cite{ulf_science}.
Since Hawking's particle production depends only on the dynamics of the probe field in the analogue geometry, the presence of an analogue black horizon is a sufficient condition to trigger the Hawking process.

\begin{figure}
\centering
  \includegraphics{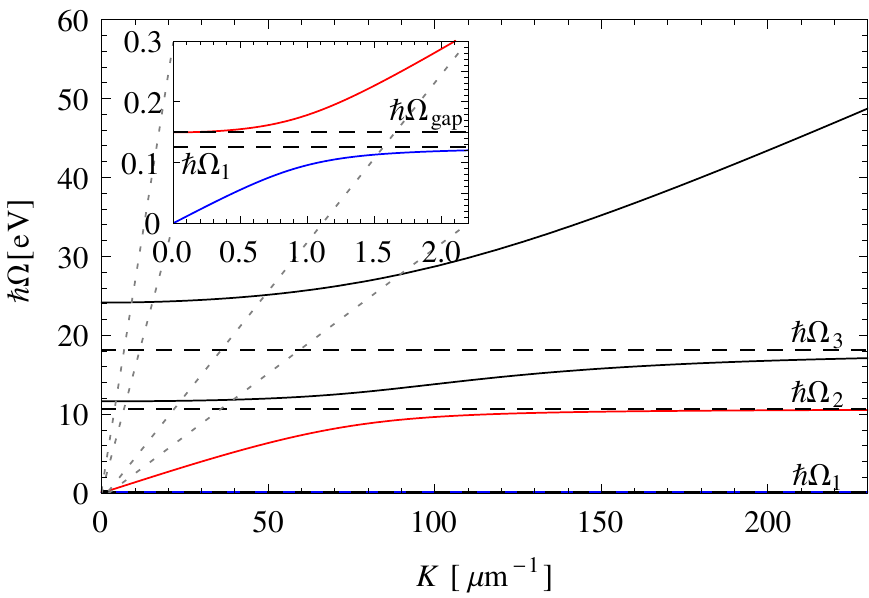}
  \hspace{3em}
  \includegraphics{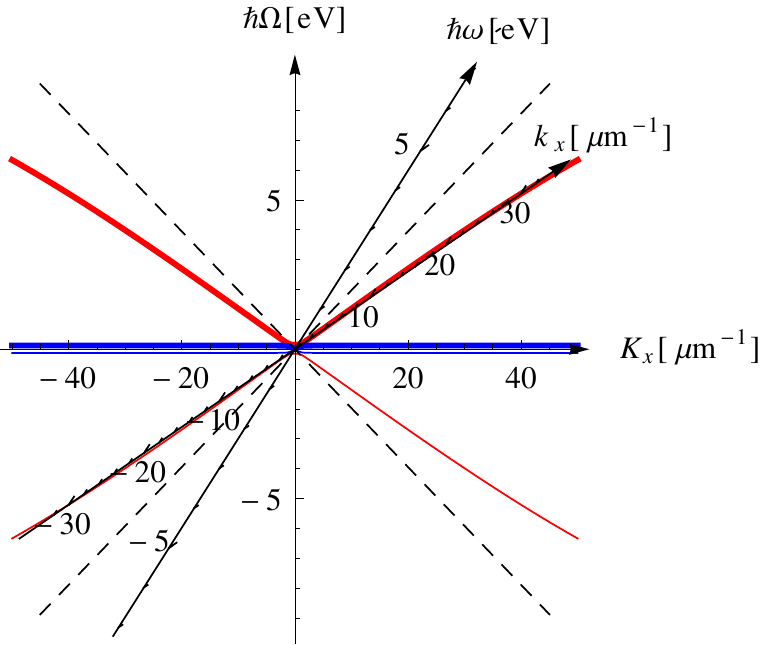}
 \caption{Left panel: Dispersion relation of fused silica~\eqref{eq:dispglass} and zoom (inset) on the small frequency region. The red line is the optical branch. The blue line, hardly visible in the main plot, is the lowermost branch. Right panel: Optical (red) and lowest (blue) branches of the dispersion relation~\eqref{eq:dispglass} in the glass rest-frame. The new axes $(\kxp,\omp)$ are obtained by applying a Lorentz transformation with $V=0.66c$ to the axes $(\kxg,\omg)$. Thick and thin lines are positive- and negative-frequency branches, respectively.}
 \label{fig:sellmeier}
\end{figure}

In the presence of dispersion, such a simple picture is more complicated. In dispersive media, indeed, quantum vacuum emission is possible even without analogue horizons, as shown in~\cite{granada_proc} in an analogue model based on a Bose-Einstein condensate.
In this section, we shall explore both horizon and horizonless configurations, where photons can be spontaneously emitted in a one-dimensional model of glass perturbed by a strong laser pulse.
For definiteness we shall work with fused silica. The dispersion relation in such a medium relates the wave number $\kg$, measured inside the glass, to the frequency $\omg$:
\begin{equation}\label{eq:dispglass}
 c^2 \kg^2=n_0^2(\omg)\omg^2,
\end{equation}
where the refractive index is given by the Sellmeier relation~\cite{refractiveindex}
\begin{equation}
 n_0^2(\omg)=1+ \sum_{i=1}^3\frac {C_i }{1-(\omg/\omg_i)^2},
\end{equation}
with
\begin{equation}
\begin{array}{ll}
 C_1=0.8974794,\qquad &\hbar\omgone=0.125285\,\mbox{eV},\\
 C_2=0.4079426,\qquad &\hbar\omgtwo=10.6661\,\mbox{eV},\\
 C_3=0.6961663,\qquad &\hbar\omgthree=18.1252\,\mbox{eV}.
\end{array}
\end{equation}
In fig.~\ref{fig:sellmeier}, left panel, we plot all the four branches of the dispersion relation, but, in what follows, the analysis is restricted to the optical branch (red line), between the upper edge of the lowest gap at $\hbar\omgap=0.149636\,\mbox{eV}$ and the next pole $\hbar\omgtwo$. We give below reasons for this choice.

To proceed, it is convenient to write the dispersion relation in the frame comoving with the pulse
\begin{equation}\label{eq:disppulse}
  c^2\left[\kyp^2+\gamma^2\left(\kxp+V\,\omp/c^2\right)^2\right]
 =\gamma^2 {\left(\omp+V\,\kxp\right)^2}\, n\!\left(\gamma\left[\omp+V\,\kxp\right],I\right)^2,
\end{equation}
by applying to eq.~\eqref{eq:dispglass} a Lorentz boost at velocity $V$
\begin{equation}\label{eq:boost}
\left\{
\begin{aligned}
  \omg &= \gamma(V)[\omp+V\,\kxp],\\
  \kxg &= \gamma(V)[\kxp+V\,\omp/c^2],\\
  \kyg &= \kyp.
\end{aligned}
\right.
\end{equation}
Hereafter, lowercase (uppercase, respectively) will denote quantities measured in the comoving (glass rest, respectively) frame.
In fig.~\ref{fig:sellmeier}, right panel, the optical branch (red) and the lowest branch (blue) of the dispersion relation are drawn in the glass rest frame $(\kxg,\omg)$.
Positive- and negative-frequency branches are plotted with thick and thin lines, respectively.
The coordinate system is then transformed under a boost with $V/c=0.66$. This value of the velocity is chosen to realize a configuration with an analogue black hole horizon as it will be shown in the following section. In practice, it can be obtained as the group velocity of a pump laser pulse with a wavelength of approximately $400~\mbox{nm}$.
In what follows, we work with frequencies much lower than $\omgtwo$, such that branches above the optical one cannot be excited. We shall therefore neglect the contribution of those branches, which are not shown in the right panel of fig.~\ref{fig:sellmeier}.

In this paper, we discuss only the kinematic properties of light propagation in the considered system. The analysis of the spectrum of the vacuum emission is postponed to further investigations~\cite{opticsgradino}.

\subsection{Horizon configurations}
\label{sec:horizons}

\begin{figure}
\centering
\includegraphics[width=\textwidth]{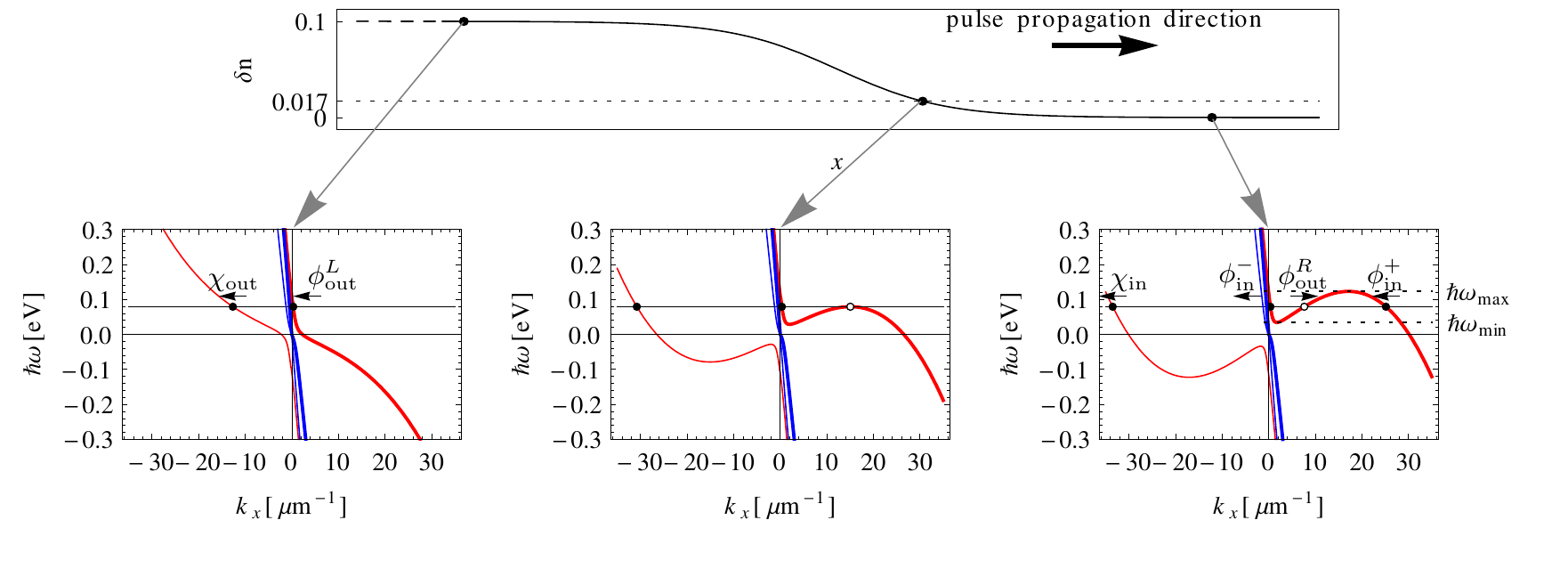}
\caption{Optical branch (red) and lowest branch (blue) of the dispersion relation~\eqref{eq:disppulse} in fused silica for vanishing transverse momentum $\kyp$, in the frame comoving with a pulse moving with velocity $V=0.66c$. Positive (negative) laboratory frequency branches are represented by thick (thin) curves.
The dispersion is plotted for three values of the perturbation of the refractive index: $\delta n=0.1$ (left panel), $\delta n=0.017$ (central panel) and $\delta n =0$ (right panel). The black horizontal line represents a generic frequency for which there are 4 real solutions in the external region (right panel), 2 real solutions in the internal one (left panel), and a point where the solutions passes from 4 to 2 (central panel). In the right panel, the dashed horizontal lines are the maximum and the minimum values of the frequency for which there are 4 solutions in this external region.
The top panel reports the positions at which the dispersion relation is computed, with respect to the leading part of the pulse, around the analogue black horizon. In the left and right panels, the arrows indicate the direction of propagations of modes corresponding to solutions on the optical branch. Modes with negative (positive) norm or, equivalently, negative (positive) laboratory frequency $\Omega$, are labeled by $\chi$ ($\phi$). Subscript in (out) refers to incoming (outgoing) modes.}
\label{fig:disp}
\end{figure}

We start by considering the configuration shown in fig.~\ref{fig:disp}. To identify the solutions of the dispersion relation for a given comoving frequency $\omp$, in the figure we plot the lowest (blue) and optical (red) branch of the dispersion relation in the frame of the pulse for the same pulse velocity $V$ as in fig.~\ref{fig:sellmeier}, right panel.
Positive (negative) laboratory frequency branches are represented by thick (thin) curves. They correspond, respectively, to positive- and negative-norm modes\footnote{A rigorous proof of this fact will be provided in~\cite{opticsgradino}, by using the norm naturally induced by the Hamiltonian structure.}.
For illustrative purposes, we choose an extremely high (possibly not so realistic) intensity $I\approx3\times10^{14}\,\mbox{W}/\mbox{cm}^2$, yielding a strong perturbation of the refractive index $\delta n=0.1$.
The dispersion relation inside the pulse (where $\delta n=0.1$) and outside the pulse (where $\delta n=0$) are plotted in the left and right panel, respectively. The top panel reports the positions where the dispersion relation is computed, with respect to the leading part of the pulse. We describe the structure of the scattering modes only around this leading edge. The discussion of the trailing edge follows the same lines.

From fig.~\ref{fig:sellmeier}, right panel, one sees that the lowest branch (blue line) saturates to $\omgone$ at small wave numbers. Except for very low comoving frequencies $\omp$ (see fig.~\ref{fig:disp}), excitations on this lowest branch have a wave number large enough that the rest-frame frequency $\omg$ is equal to $\omgone$, independently of the momentum $\kxg$.
As a consequence, these particles cannot propagate, because they have zero group velocity in the glass rest frame. Eventually they are completely absorbed.
Thus, we shall consider only the optical branch. The case in which $\omp$ is so small that the lowest branch cannot be neglected requires a different analysis, beyond the scope of this paper.

We now introduce the concept of frequency-dependent horizon. Assuming that the pulse is moving in the positive $x$ direction ($V>0$), such a horizon is present for a frequency $\omp$ when inside the pulse there are only negative group velocity (measured in the comoving frame) modes, while outside there are both negative and positive group velocity modes.
In this case, light can propagate only left-ward inside the pulse, both left- and right-ward outside it.
Equivalently, there exists a point in between the internal and the external regions such that the speed of light in the comoving frame vanishes at this frequency,
\begin{equation}\label{eq:zerovel}
 \frac{\dd\omp}{\dd \kxp}=0.
\end{equation}
In fig.~\ref{fig:disp}, the dispersion relation is graphically solved on the optical branch (red curves) for a generic frequency (solid horizontal line). The arrows indicate the direction of propagation of the corresponding modes, which are labeled in the left and right panel by $\chi$ ($\phi$) if they have negative (positive) laboratory frequency $\Omega$, or, equivalently, negative (positive) norm. The subscript in (out) denotes incoming (outgoing) modes, having group velocity directed toward (away from) the horizon.
Note how the nontrivial shape of the dispersion relation~\eqref{eq:dispglass} is responsible for new interesting features with respect to the standard models of subluminal dispersion usually considered in the literature~\cite{lr}.

For the chosen value of the comoving frequency $\om$, eq.~\eqref{eq:zerovel} is satisfied at a point where $\delta n\approx0.017$. For this particular value of $\delta n$, the dispersion relation (central panel) is tangent to the line of constant $\omp$ (two coincident solutions denoted by an empty dot). For outer points (right panel, $\delta n<0.017$), the dispersion relation has 4 solutions. Three of them correspond to incoming modes ($\chi_{\rm in}$, $\phi_{\rm in}^-$, and $\phi_{\rm in}^+$), since they propagate from $x=+\infty$ toward the horizon, having negative group velocity.
The fourth one (denoted by an empty dot) corresponds instead to an outgoing mode ($\phi_{\rm out}^R$), with positive group velocity.
For inner points (left panel, $\delta n>0.017$) the dispersion relation has only 2 real solutions, both corresponding to outgoing modes ($\chi_{\rm out}$ and $\phi_{\rm out}^L$), which, having negative group velocity, propagate from the horizon toward $x=-\infty$.
As a result, according to the definition given above eq.~\eqref{eq:zerovel}, within the frequency range $\ommin<\om<\ommax$ (where $\hbar\ommin\approx0.034\,\mbox{eV}$ and $\hbar\ommax\approx0.123\,\mbox{eV}$ are represented in the right panel with dotted lines), the leading edge behaves as a black hole horizon. Analogously, the trailing edge behaves as a white hole horizon. In a more realistic situation, for typical experimental parameters~\cite{faccioexp}, one has a much smaller $\delta n$.
As a result obtaining a configuration with horizon requires a very accurate tuning of the pulse velocity, down to the fourth digit.
For instance, with $\delta n\approx 0.001$, a suitable value of the pulse velocity is $v\approx0.6838c$. Furthermore, the horizon turns out to be present only in a very narrow frequency window, between $\hbar\ommin\approx0.01379\,\mbox{eV}$ and $\hbar\ommax\approx0.01396\,\mbox{eV}$, for the considered parameters.

\begin{figure}
\centering
 \includegraphics{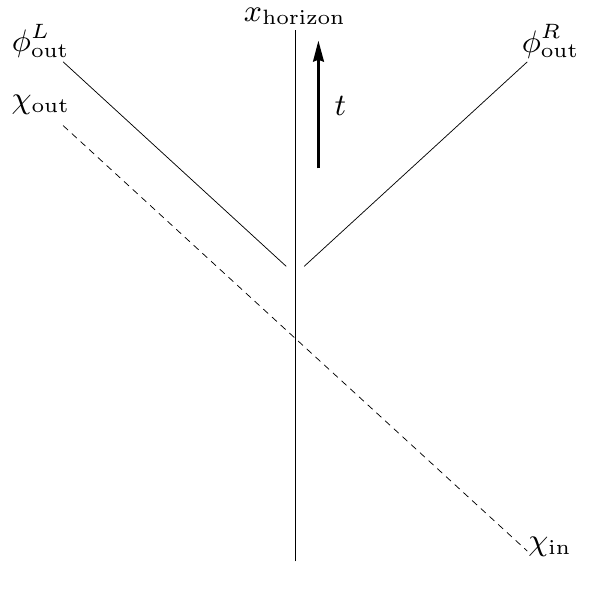}
 \caption{\label{fig:decomposition}Decomposition of the incoming mode $\chi_{\rm in}$ into outgoing modes. This mode travels from $x=+\infty$ toward the horizon where it is scattered in the three outgoing modes $\chi_{\rm out}$, $\phi_{\rm out}^L$, and $\phi_{\rm out}^R$. Positive (negative) norm modes are represented by solid (dashed) lines.}
\end{figure}

Summarizing, for all frequencies $\ommin<\omp<\ommax$, there are 3 incoming ($\chi_{\rm in}$, $\phi_{\rm in}^-$, and $\phi_{\rm in}^+$) and 3 outgoing ($\chi_{\rm out}$, $\phi_{\rm out}^L$, and $\phi_{\rm out}^R$) modes. Consequently, the scattering (conversion of incoming modes into outgoing ones) at the analogue black horizon is described by a $3\times3$ matrix~\cite{MacherRP1}.
For instance, if one sent a negative-norm wave $\chi_{\rm in}$ into the system, it would be decomposed in a superposition of the three outgoing modes (see fig.~\ref{fig:decomposition}),
\begin{equation}
\chi_{\rm in}\longrightarrow \alpha\chi_{\rm out}+B\phi_{\rm out}^L+\beta\phi_{\rm out}^R.
\end{equation}
The three coefficients of this decomposition form a row of the scattering matrix. Analogously, the remaining two rows are obtained by decomposing the other two incoming modes. Since the scattering matrix mixes modes with positive and negative norm/laboratory frequency, the initial vacuum (i.e., the quantum state with no incoming particles) is not equivalent to the final vacuum. As usual, this implies that quanta on outgoing modes can be spontaneously emitted starting from initial vacuum fluctuations. In particular, it is easy to show that the occupation number of the positive-norm outgoing mode $\phi_{\rm out}^R$ is given by $|\beta|^2$.
In a different analogue model based on flowing atomic condensates~\cite{garay}, scattering is also described by a $3\times3$ matrix and the spectrum of spontaneously emitted particles, given by $|\beta|^2$, follows a Planckian distribution, as for standard Hawking radiation, but only up to a certain frequency $\ommax$~\cite{MacherBEC,Recati2009,iacopo}.
Shall we observe Hawking-like radiation also in the present system?
Unfortunately, the scattering is now described by a $3\times3$ matrix only for $\omp>\ommin$. In fact, the two modes $\phi_{\rm out}^R$ and $\phi_{\rm in}^+$ disappear for $\omp<\ommin$, thus the scattering matrix becomes $2\times2$ and $\beta$ is not defined.
However, since a negative frequency incoming mode ($\chi_{\rm in}$) is still present, quantum vacuum emission is expected to appear also in this case, albeit with different spectral properties.

\subsection{Horizonless configurations: fast pulse}

%
\begin{figure}[b]
 \centering
 \includegraphics{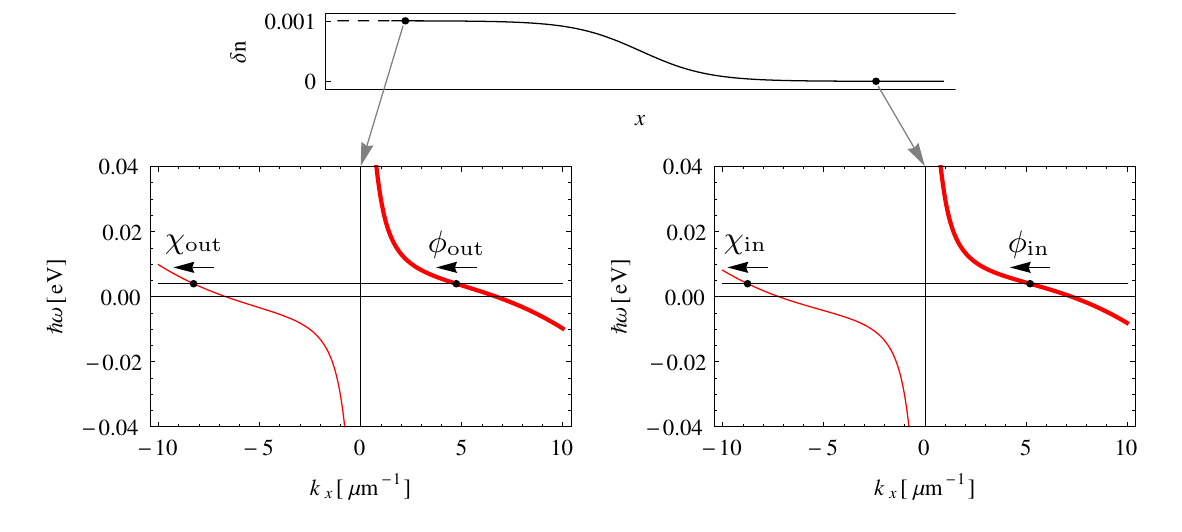}
  \caption{Positive (thick curves) and negative (thin curves) norm/laboratory frequency branches of the dispersion relation~\eqref{eq:disppulse} in fused silica for vanishing transverse momentum $\kyp$, in the frame comoving with a pulse moving with velocity $V\approx0.69c$. $\delta n=0.001$ (left panel) and $\delta n=0$ (right panel) as in the experiment of~\cite{faccioexp,faccioexplong}. In both the internal (left) and external (right) regions only two real solutions are present.
  Arrows indicate the direction of propagations of modes. Modes with negative (positive) norm or, equivalently, negative (positive) laboratory frequency $\Omega$, are labeled by $\chi$ ($\phi$). Subscript in (out) refers to incoming (outgoing) modes.}
  \label{fig:dispfaccio}
\end{figure}
As a second case, we consider a pulse propagating with a large velocity $V$, 
such that the probe group velocity $v$ is smaller than $V$ for any frequency $\omp$. In this situation, no positive group velocity modes propagate either inside or outside the pulse, {\it i.e.} condition~\eqref{eq:zerovel} is not satisfied.
This configuration was experimentally realized in~\cite{faccioexp,faccioexplong} and theoretically investigated in~\cite{faccioprl}, where the nonlinearity of the dispersion relation is neglected in the quantization of the probe field. Dispersive effects have also been taken into account in a recent analysis~\cite{facciopra}.
In fig.~\ref{fig:dispfaccio}, we plot the dispersion relation inside the pulse (left panel) and outside (right panel) for this experimental set up ($V=0.69c$, $\delta n=0.001$).
Thick (thin) lines denote positive (negative) norm/laboratory frequency branches. With the same convention introduced in the above section, modes with negative (positive) norm are named $\chi$ ($\phi$).
The dispersion relation has now only two real solutions on both sides. All of them correspond to leftgoing modes, two of them correspond to incoming modes $\chi_{\rm in}$ and $\phi_{\rm out}$, and two of them to outgoing modes $\chi_{\rm out}$ and $\phi_{\rm out}$.
Since the dispersion relation has now only two real solutions, the scattering process is described by a $2\times2$ matrix. An incoming negative laboratory frequency mode $\chi_{\rm in}$ is still present and its decomposition in the two outgoing modes is
\begin{equation}
 \chi_{\rm in}\longrightarrow \alpha\chi_{\rm out}+B\phi_{\rm out}.
\end{equation}
The disappearance of the horizon and, consequently, of the outgoing mode $\phi_{\rm out}^R$ removes the Hawking-like channel of particle production.
Nevertheless, particles can be spontaneously emitted because the scattering matrix still mixes positive- and negative-norm modes. For instance, the occupation number of the outgoing mode $\phi_{\rm out}$ is given by $|B|^2$.

\subsection{Horizonless configurations: slow pulse}
As the last case we consider a pulse which is slow enough that, for some frequencies, the group velocity $v$ of the probe field is larger than the pulse velocity $V$, both inside and outside the pulse. This implies that no horizon is present for those frequencies.
Nevertherless, for all values of $\omp$, negative-frequency modes are present, so that quantum vacuum emission is expected. This is possible because the glass rest frame frequency $\omg$ goes to a finite value $\omgtwo$ for $k\to\infty$.

\begin{figure}
 \centering
 \includegraphics{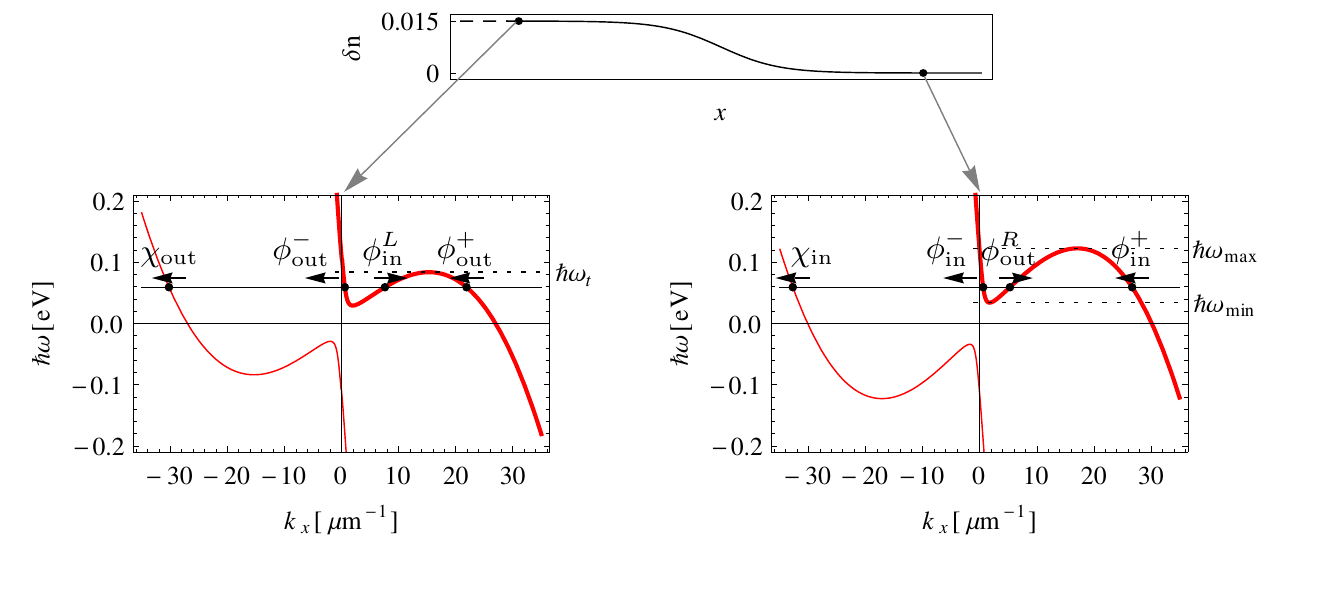}
  \caption{Positive (thick curves) and negative (thin curves) norm/laboratory frequency branches of the dispersion relation~\eqref{eq:disppulse} in fused silica for vanishing transverse momentum $\kyp$, in the frame comoving with a pulse moving with velocity $V\approx0.66c$. $\delta n=0.015$ (left panel) and $\delta n=0$ (right panel).
  There exists a threshold frequency $\omlow$ (left panel) such that, for $\omlow<\om<\ommax$, the number of propagating modes is 4 in the external region (right) and 2 in the internal one (left), as in the system of fig.~\ref{fig:disp}. However, for $\om<\omp<\omlow$ (horizontal solid line), the number of modes is 4 in both regions, similar to what was found in~\cite{granada_proc,warpdrive} for superluminal (supersonic, for analogue BEC systems) systems with superluminal (supersonic) dispersion relation. Finally, for $\om<\ommin$ or $\om>\ommax$, there are two real solutions of the dispersion relation.
  Arrows indicate the direction of propagations of modes. Modes with negative (positive) norm or, equivalently, negative (positive) laboratory frequency $\Omega$, are labeled by $\chi$ ($\phi$). Subscript in (out) refers to incoming (outgoing) modes.}
  \label{fig:dispslow}
\end{figure}

In this configuration, the dispersion relation (fig.~\ref{fig:dispslow}) has now a maximum both in the external ($\delta n=0$, right panel) and in the internal region ($\delta n=0.015$, left panel), for a pulse velocity $V=0.66c$. Consequently, there is a threshold frequency $\omlow$ (left panel), such that, for $\omlow<\omp<\ommax$, the system is described by a $3\times3$ scattering process as in configurations with horizons.
For $\omp<\ommin$ and for $\omp>\ommax$, there are instead only 2 real solutions in both regions, as in the above discussed case.
Finally, for $\ommin<\omp<\omlow$ (horizontal solid line), there are 4 real solutions in both regions. 
They correspond to 4 incoming modes, with group velocity directed toward the horizon ($\phi_{\rm in}^L$, propagating rightward in the left region; $\chi_{\rm in}$, $\phi_{\rm in}^-$, and $\phi_{\rm in}^+$, propagating leftward in the right region), and 4 outgoing modes, with group velocity directed away from the horizon  ($\chi_{\rm out}$, $\phi_{\rm out}^-$, and $\phi_{\rm out}^+$, propagating leftward in the left region; $\phi_{\rm in}^R$, propagating rightward in the right region). Two ($\chi_{\rm in}$ and $\chi_{\rm out}$) of these 8 modes have negative norm/laboratory frequency.
As a consequence, the scattering is now described by a $4\times4$ matrix, and the decomposition of the incoming negative-frequency mode is
\begin{equation}
\chi_{\rm in}\longrightarrow \alpha\chi_{\rm out}+B^-\phi_{\rm out}^-+B^+\phi_{\rm out}^++\beta\phi_{\rm out}^R.
\end{equation}
From this expression one can read the occupation numbers $|B^-|^2$, $|B^+|^2$, and $|\beta|^2$ of the positive-norm outgoing modes $\phi_{\rm out}^-$, $\phi_{\rm out}^+$, and $\phi_{\rm out}^R$, computed on the initial vacuum of the incoming modes.
A similar situation was studied in~\cite{granada_proc,warpdrive} in a Bose-Einstein condensate, where the dispersion relation is supersonic.
In that case it has been shown that radiation is spontaneously emitted, but the spectrum is not Planckian.
We expect that those results can be directly extended to the present situation since, for a given comoving frequency $\om$, a system with subluminal dispersion relation where $v>V$ is equivalent to a system with superluminal dispersion relation where $v<V$~\cite{cpf}.

\section{Propagation in three-dimensional systems}

In the above section we described the configurations where quantum vacuum emission may be present in an optical system.
For the sake of simplicity we restricted our analysis to a one-dimensional model, with light propagating on the axis parallel to the pulse velocity.
Focusing on the configuration with analogue horizons (sec.~\ref{sec:horizons}),
we now relate the features of the emission in the comoving frame to what can be detected in the laboratory, outside the glass.
We therefore consider the propagation of emitted photons in a three-dimensional system with axial symmetry, thus effectively two-dimensional.
Moreover, we assume that the direction $x$ of propagation of the pulse is parallel to the interface between glass and air, through which observations are made, as in~\cite{faccioexp}. The $y$-axis is then orthogonal to this interface. For the sake of simplicity, the air refractive index is taken equal to 1.
Frequencies and wave numbers will be denoted by the subscript ``air'' when measured outside the glass. According to the notation adopted in the previous section no subscript indicates quantities measured in the glass, uppercase refers to the glass rest frame, lowercase to the frame comoving with the pulse at $V\approx0.66c$ as in fig.~\ref{fig:disp}.

\begin{figure}
\includegraphics[width=\textwidth]{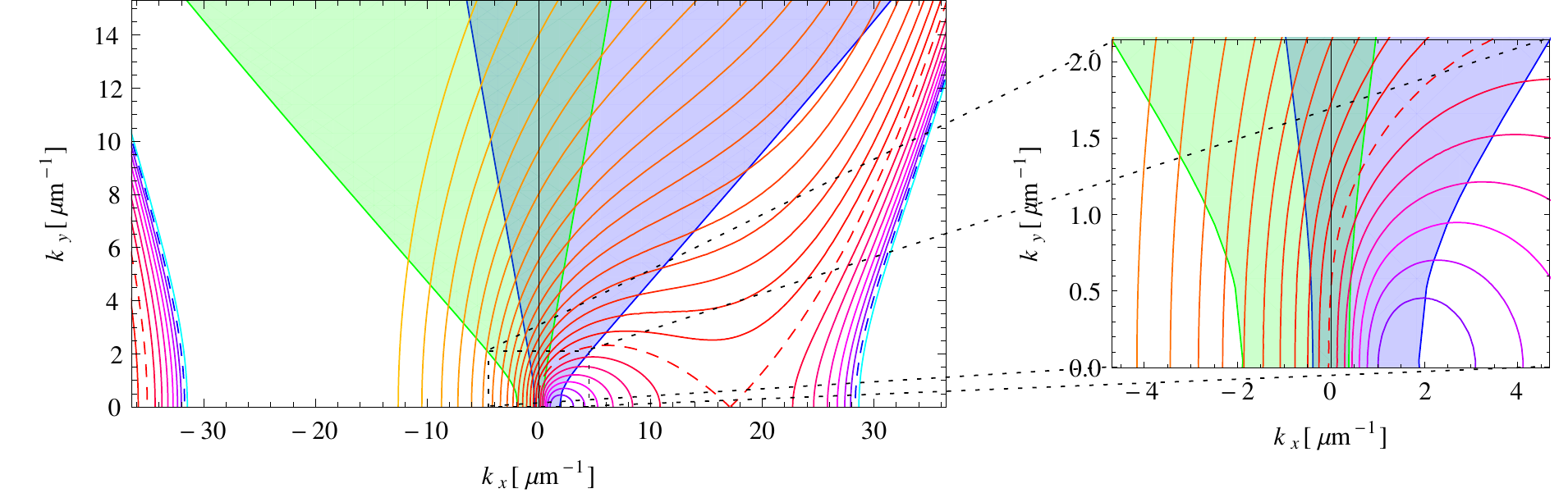}
\caption{\emph{Loci} of constant emission frequency $\omp$ (solid lines), equally spaced in $\log(\omp)$, in the comoving frame ($\kxp,\kyp$), with $V=0.66c$, inside the glass.
The dashed lines are $\omp=\ommin$ (blue), and $\omp=\ommax$ (red). The cyan curve refers to a frequency $\om<\ommin$, from blue to red to $\ommin<\om<\ommax$, and from red to yellow to $\om>\ommax$.
In the right panel the low momentum region is zoomed.
A mode with positive (negative) norm/laboratory frequency does not undergo total reflection and can cross the interface between glass and air, if the corresponding wavevector ($\kxp$,$\kyp$) is contained in the green (blue) area.}
\label{fig:paperocomoving}
\end{figure}
We begin by studying the properties of the dispersion relation inside the glass, but outside the pulse (corresponding to the right panel of fig.~\ref{fig:disp}).
In fig.~\ref{fig:paperocomoving}, \emph{loci} of constant $\omp$ are plotted in the comoving frame $(\kxp,\kyp)$.
The implicit relation between $\kxp$ and $\kyp$, in the comoving frame, at fixed $\omp$, is directly provided by eq.~\eqref{eq:disppulse}.
In the right panel the low momentum region is zoomed.
The curves at fixed $\omp=\ommin$ and $\omp=\ommax$ are represented by blue and red, respectively, dashed lines.

To explain the meaning and relevance of this plot, we first compare the section at $\kyp=0$ with the dispersion relation of fig.~\ref{fig:disp}, right panel.
For $\ommin<\omp<\ommax$ (colors from blue to red), there are four real solutions for $\kxp$, at fixed $\om$, represented by a single color between blue and red, and $\kyp=0$ (cf. the 4 solutions on the solid horizontal line of constant $\omp$, in fig.~\ref{fig:disp}).
Then, when the frequency increases, the two rightmost solutions merge into a single one at $\omp=\ommax$ (red dashed line) and disappear for $\omp>\ommax$ (colors from red to yellow). When, instead, the frequency goes down to $\omp=\ommin$ (blue dashed line), the two central solutions collapse into a single point, around $\kxp\approx 2\mu\mbox{m}^{-1}$.
Finally, for $\omp<\ommin$ (cyan curve), they completely disappear.
In the full two-dimensional case, {\it i.e.}, when $\kyp\neq0$, the situation is more complicated and four solutions for $\kxp$ may exist even for $\omp>\ommax$, for sufficiently small but nonzero values of $\kyp$ (see, for instance, the first red curve above the dashed red one).

Noticeably, for all values of $\omp$, there is always one solution on the leftmost branch.
By comparison with fig.~\ref{fig:disp}, right panel, this solution has negative rest frame frequency $\omg$. The leftmost solution in fig.~\ref{fig:disp} lies indeed on the negative laboratory frequency branch, represented therein by a thin line.
As noticed in the previous section, some quantum vacuum emission should therefore be present at any frequency $\omp$.

However, in order to be detected, phonons must be able to travel outside the glass reaching a detector at rest in the laboratory.
To determine which photons can really reach the detector and what is their frequency and momentum in the laboratory, one must first make a Lorentz boost to the glass rest frame and then study how light exits the glass.

\begin{figure}
\includegraphics[width=\textwidth]{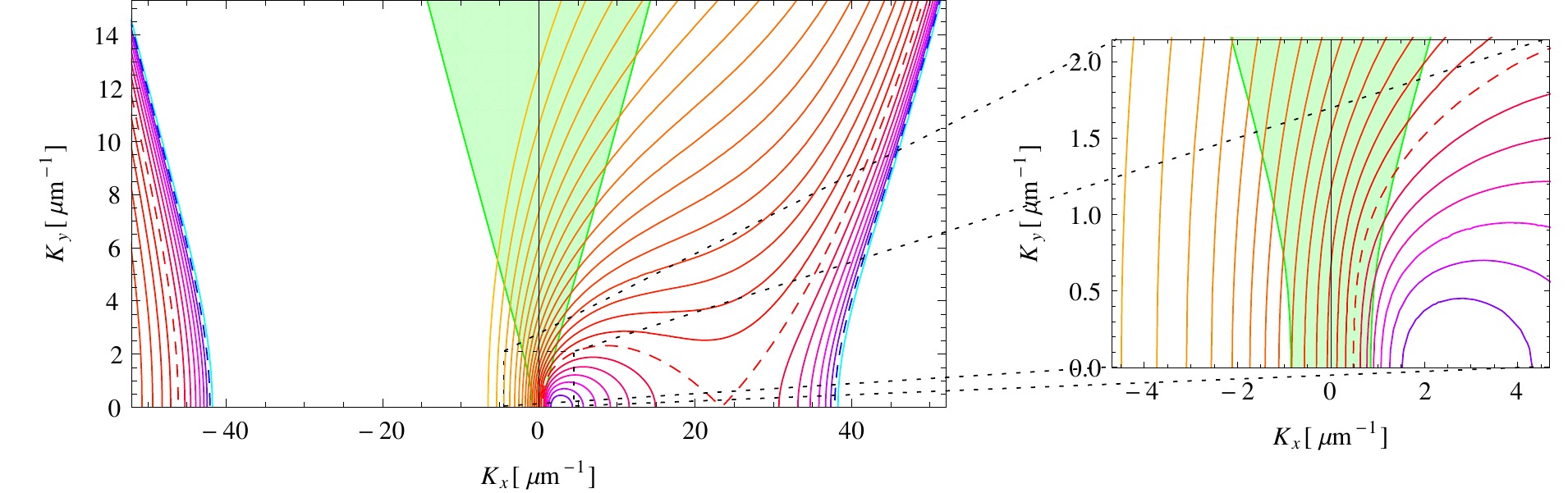}
\caption{\emph{Loci} of constant emission frequency $\omp$ (solid lines), equally spaced in $\log(\omp)$, in the glass rest frame ($\kxg,\kyg$), inside the glass.
The dashed lines are $\omp=\ommin$ (blue), and $\omp=\ommax$ (red). Colors from cyan to blue refer to frequency $\om<\ommin$, from blue to red to $\ommin<\om<\ommax$, and from red to yellow to $\om>\ommax$.
In the right panel the low momentum region is zoomed.
A mode does not undergo total reflection and can cross the interface between glass and air, if the corresponding wavevector ($\kxg$,$\kyg$) is contained in the green area.}
\label{fig:paperoglass}
\end{figure}

Thus, in fig.~\ref{fig:paperoglass} we report, in the rest frame of the glass ($\kxg$, $\kyg$), the curves of fixed comoving frequency for the same values of $\omp$ as fig.~\ref{fig:paperocomoving}, maintaining the same color correspondence.
The relation between $\kxg$ and $\kyg$, at fixed comoving frequency $\omp$, is obtained by solving
\begin{equation}
 \omp = \gamma(V)[\omg-V\,\kxg]
\end{equation}
for $\omg$ and putting the result into eq.~\eqref{eq:dispglass}:
\begin{equation}
 c^2(\kyg^2+\kxg^2)=\left(\omp/\gamma+v\,\kxg\right)^2 n_0^2\!\left(\omp/\gamma+v\,\kxg\right).
\end{equation}

It is now possible to study whether a mode with a given $\omp$, $\kxg$ and $\kyg$ can reach or not a detector placed outside the glass, by crossing the interface between glass and air. In fig.~\ref{fig:paperoglass}, the green area corresponds to rays that do not undergo total reflection and exit the parallelepiped through the lateral surface.
Its boundary is the set of wave vectors $(\kxg,\kyg)$ determined by imposing that $\ky$ vanishes in air,
\begin{equation}\label{eq:omtotal}
 \oml^2=c^2\kx^2.
\end{equation}
Since $\oml=\omg$ and $\kx=\kxg$, from eq.~\eqref{eq:disppulse} we obtain
\begin{equation}\label{eq:totalg}
 \kyg^2=\kxg^2\left[n_0^2(c\kxg)-1\right].
\end{equation}
For the sake of completeness, we report in fig.~\ref{fig:paperocomoving} the corresponding locus of wavevectors describing photons that can cross the lateral surface of the parallelepiped in the comoving frame $(\kxp,\kyp)$.
The procedure to determine this locus is analogous to the above one.
However, the result depends on the sign of the rest frame frequency $\omg$ of the considered mode.
The green area in fig.~\ref{fig:paperoglass} is transformed into the green or blue area in fig.~\ref{fig:paperocomoving}, when $\omg>0$ or $\omg<0$, respectively. 
That is, if a mode has positive (negative) norm/laboratory frequency, then it does not undergo total reflection if the corresponding wave vector ($\kxp,\kyp$) lies in the green (blue) area of fig.~\ref{fig:paperocomoving}.

It is now possible, either using figs.~\ref{fig:paperocomoving} or~\ref{fig:paperoglass} to determine which modes can be observed outside the glass.
Solutions on the leftmost and rightmost branches all lie in the white area, thus the corresponding modes cannot cross the interface parallel to the direction of the beam.
Consequently, modes with $\omp<\ommin$, whose wave number lies on those branches (cyan curve), cannot be detected in the laboratory through the lateral surface%
\footnote{Of course, if the glass has defects, photons may be elastically scattered and exit the parallelepiped from the lateral surface, even if the $x$ component of their momentum is initially large.}.
We then consider modes with $\ommin<\omp<\ommax$ (from blue to red curves). As shown in sect.~\ref{sec:horizons}, this is the frequency range where the scattering is described in the standard way by a $3\times3$ matrix. In this range, there are solutions with low momentum $k$, therefore quantum vacuum emission is expected to be stronger, by momentum conservation arguments.
Unfortunately, only a tiny region of wave vectors with comoving frequency $\ommin<\omp<\ommax$ is contained in the green area.
Finally, the largest part of the green area in the central panels is filled by modes with $\omp>\ommax$ (from red to yellow curves). Those photons, with very high comoving energy can cross the lateral surface and may be observed in the laboratory.

After having determined which modes can exit the glass, we still have to understand how they look like in the laboratory, what is their wave number, and, in particular, what is their propagation direction.
\begin{figure}
\centering
\includegraphics[width=0.58\textwidth]{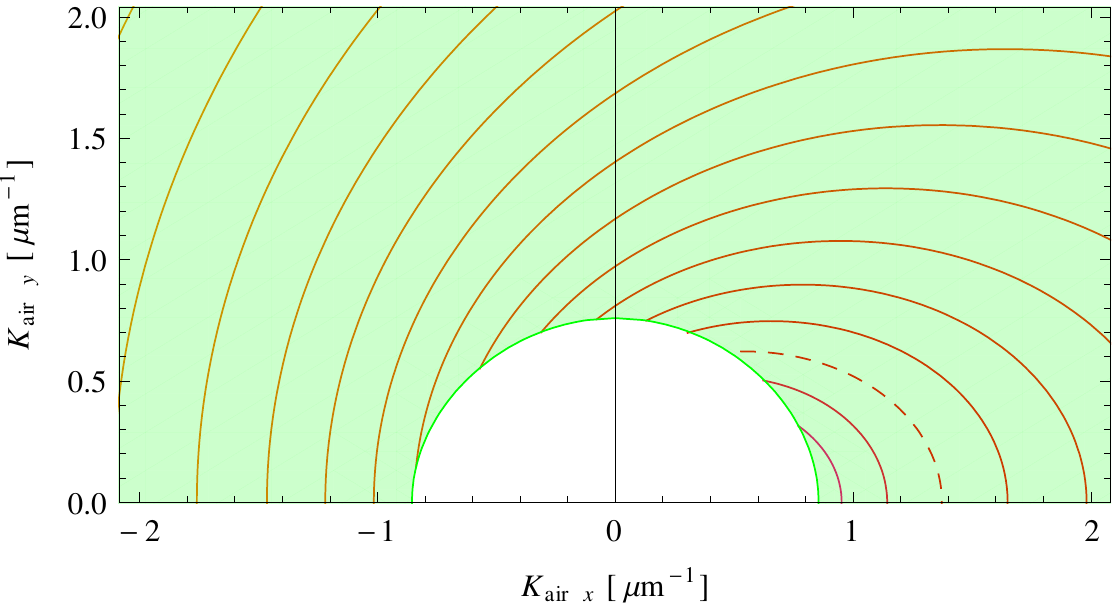}
\caption{\emph{Loci} of constant emission frequency $\omp$ (solid lines), equally spaced in $\log(\omp)$, as observed in the laboratory reference frame where the glass is at rest ($\kx,\ky$), outside the glass.
The red dashed line is $\omp=\ommax$. Colors from blue to red refer to $\ommin<\om<\ommax$, and from red to yellow to $\om>\ommax$.
A mode can originate on the optical branch inside the glass if the corresponding wavevector ($\kx$,$\ky$) is contained in the green area.}
\label{fig:paperolab}
\end{figure}
In fig.~\ref{fig:paperolab}, we plot the \emph{loci} of constant $\omp$ in air, in the glass rest frame $(\kx,\ky)$, for the same values of $\omp$ as in figs.~\ref{fig:paperocomoving} and~\ref{fig:paperoglass}, with the same color legend.
Since in the passage from glass to air, both the frequency and the $x$ component of the wave vector are conserved, the dispersion relation in the comoving frame in air is
\begin{equation}
 c^2(\kxb^2+\kyb^2)=\omb^2=\omp^2.
\end{equation}
To determine the \emph{loci} of constant $\omp$ in the laboratory reference frame in air, one must apply a Lorentz boost to this dispersion relation. However since this equation describes circles in the plane $(\kxb,\kyb)$ and a Lorentz transformation is affine, then \emph{loci} of constant $\omp$ will appear as ellipses in the plane $(\kx,\ky)$.
Modes having wavevector in the green area can propagate inside the glass on the optical branch. The white area corresponds instead to modes that either do not belong to the optical branch ($\oml=\omg<\omgap$) or are evanescent waves in glass.

To conclude this section, we compare the \emph{loci} of constant $\omp$ in the glass rest frame inside and outside the glass.
The segments of curves lying in the green area of fig.~\ref{fig:paperoglass} (rest frame in glass) are mapped onto the whole curves in fig.~\ref{fig:paperolab} (rest frame in air).
Instead, the segments of curves lying in the white area of fig.~\ref{fig:paperoglass} do not have any counterpart in fig.~\ref{fig:paperolab} because of total reflection. That is, by looking through the interface parallel to the direction of propagation of the pulse, it is not possible to investigate the very interesting mode structure present inside the glass.

Note that, in the laboratory frame, outside the glass, only a narrow region of wave vectors $(\kx,\ky)$ stays below the red dashed curve ($\omp=\ommax$), then corresponding to modes with comoving frequency smaller than $\ommax$. In particular, no modes with $\om<\ommax$ (from blue to red curves) propagate in air perpendicularly to the direction of the beam ($\ky=0$). Furthermore the blue dashed curve and the cyan curve of figs.~\ref{fig:paperocomoving} and~\ref{fig:paperoglass} do not have any counterpart in fig.~\ref{fig:paperoglass}. This is not surprising because, as we already noticed, the blue dashed curve is not contained in the green area of figs.~\ref{fig:paperocomoving} and~\ref{fig:paperoglass}, implying that modes with comoving frequency $\omp<\ommin$ cannot propagate in air, by crossing the lateral surface of the parallelepiped.
We also noticed that the majority of modes that cross the lateral surface have frequency $\om>\ommax$ (from red to yellow curves). A lot of them even propagate in the direction perpendicular to the beam, but unfortunately they cannot be photons produced in a standard analogue Hawking mechanism, because their comoving frequency $\omp$ is too high.

\section{Constraints from observations}

Here, we reverse the logic of the previous section, by determining the constraints put on the emission properties by a given observation in the laboratory.
Let us assume that some observation is performed in the laboratory in a frequency range $(\omlone,\omltwo)$.
What is the frequency window of emission $\omp$ compatible with those measures?
The analysis is straightforward and the result is independent of the nature of the material.
Since frequency is conserved when light crosses the interface ($\omg=\oml$, $\omp=\omb$), radiation observed at frequency $\oml$ in the laboratory must have been emitted at
\begin{equation}
 \omp=\gamma(V)\left[\oml-V\,\kx\right].
\end{equation}
Assuming again that the air refractive index is 1, one obtains
\begin{equation}
 \omp=\oml\,\gamma(V)\left[1-V\,\cos\thetal/c\right],
\end{equation}
where $\thetal$ is the angle of propagation in air with respect to the pulse direction.
Letting $\cos\thetal$ vary from $-1$ to $+1$, the emission frequency window in the comoving frame is
\begin{equation}
 \left(\ompone=\omlone\gamma(V)\left[1-V/c\right],\omptwo=\omltwo\gamma(V)\left[1+V/c\right]\right).
\end{equation}
If measurements are performed by an instrument with frequency window $(\omlone,\omltwo)$, only emission frequencies in the range $(\ompone,\omptwo)$ can be explored. Using a pulse velocity of $0.7 c$, approximately typical of fused silica, $\ompone=0.4\,\omlone$ and $\omptwo=2.4\,\omltwo$. In particular, photons produced in the comoving frame at low $\omp$ cannot be detected.

Conversely, let us assume that measurements are performed through a surface {\it perpendicular} to the direction of propagation of the pulse.
In this case, radiation forming inside the glass a very small angle $\thetagl$ with respect to the direction of the pulse impinges almost perpendicularly on the new interface.
This radiation does not undergo total reflection and can be detected.
The emission frequency $\om$ in the comoving frame is then expressed in term of the glass rest frame frequency $\omg=\oml$ as
\begin{equation}\label{eq:ompparallel}
 \omp=\gamma(V)\left[\omg-V\,\kxg\right]
 =\gamma(V)\left[\omg\mp V\,\sqrt{n_0^2(\omg)\omg^2/c^2-\kyg^2}\right]
 =\gamma(V)\oml\left[1\mp \frac{V}{c}\,\sqrt{n_0^2(\oml)-\sin^2\!\thetal}\right],
\end{equation}
since the $y$ component of the momentum is now conserved in crossing the interface. The upper (lower) sign applies to photons observed through the front (rear) interface.
Since the group velocity of the pulse is not very different from the phase velocity at the pump frequency $\omlaser$, the pulse velocity is $V\approx c/n_0(\omlaser)$. As a consequence, if $\omlaser$ is not far from the observation window $(\omlone,\omltwo)$, $n_0(\oml)\approx n_0(\omlaser)\approx c/V$, and eq.~\eqref{eq:ompparallel} becomes
\begin{equation}
 \omp\approx\gamma(V)\oml\left[1\mp \sqrt{1-\frac{V^2}{c^2}\sin^2\!\thetal}\right],
\end{equation}
If measurements are performed through the front interface ($0<\thetagl<90^\circ$, minus sign in the parenthesis), the emission window in the comoving frame compatible with observations in the range $(\omlone,\omltwo)$ is then $\left(0,(\gamma-1)\omltwo\right)$.
Interestingly, it goes down to very low frequencies, even if the observation window does not. This result implies that 
using a detector in a given frequency range ($\omlone,\omltwo$), one can explore comoving emission frequencies in a very low frequency range.

If measurements are performed through the rear interface ($90^\circ<\thetagl<180^\circ$, plus sign in the parenthesis) the emission window compatible with observations in the same frequency range ($\omlone,\omltwo$) lies instead at much higher frequencies $\left((\gamma+1)\omlone,2\gamma\omltwo\right)$.

\section{Conclusions}

In the first part of the paper, we used simple kinematic arguments to identify under what conditions quantum vacuum radiation can be generated from a strong laser pulse propagating in a nonlinear Kerr medium such as fused silica. We identified three possible configurations:

1) \emph{Horizon geometry}. Even if the medium dispersion hinders the usual analogy with the propagation of a quantum field on a curved spacetime, frequency-dependent horizons can still be defined in a finite comoving frequency range $\ommin<\om<\ommax$. Seen by a comoving observer, light within this frequency window can propagate only in one direction inside the pulse, while it propagates in both directions in the external region. In analogy with other analogue models based, {\it e.g.}, on Bose-Einstein condensates, we expect that Hawking-like thermal radiation should appear for frequencies $\om$ far enough from $\ommin$ and $\ommax$. 

2) \emph{Horizonless geometry, fast pulse}. If the pulse is fast enough,  in the comoving frame modes can propagate only in one direction, both inside and outside the pulse. However, as negative-frequency modes exist also in this configuration, quantum vacuum emission is possible, but its spectrum is expected to strongly deviate from a Planckian law.

3) \emph{Horizonless geometry, slow pulse}. If the pulse is too slow, modes at frequencies $\omega$ below a threshold frequency $\om_t$ can propagate in the comoving frame in both directions both inside and outside the pulse, but a frequency-dependent horizon is present for $\om_t<\om<\ommax$. Sufficiently far from $\om_t$ and $\ommax$, the emission is then expected to be Hawking-like as in 1). For all other frequencies, quantum radiation is still possible, yet with a different spectrum.

On this basis, we are tempted to conclude that configuration 1) is the most suitable to observe the optical analogue of Hawking radiation.

In the second part of the paper, we focus on configuration 1) to describe the propagation of quantum radiation from the pulse inside the nonlinear medium (where it is emitted) to the detector that is located in the surrounding air. In particular, we identified those modes that can cross the interface between glass and air. This analysis yields two important results: first, the Hawking-like emission within the $\ommin<\om<\ommax$ window does not propagate outside the glass perpendicularly to the direction of propagation of the pulse. Second, the dispersive high momentum modes cannot cross a glass/air interface parallel to the pulse direction. This implies that the Hawking partners cannot be detected looking through this surface either.

In conclusion, these kinematic arguments show that the optimal directions to observe Hawking-like emission in a realistic experimental setup as in~\cite{faccioexp} form a very narrow angle with the direction of the pulse. In addition to this, quantum radiation is anticipated to appear also in horizonless configurations. However, its spectral features are expected to strongly differ from those of standard analogue Hawking emission.

\begin{acknowledgement}
The authors thank D. Faccio for a careful reading of this work. SF thanks A. Coutant, S. Liberati, A. Prain, and R. Parentani for stimulating discussions.
\end{acknowledgement}

\bibliography{kinetics}
\bibliographystyle{epjp}

\end{document}